\documentstyle[11pt,aaspp4]{article}
\begin{document}  
\title{THE ROTATION VELOCITY - DENSITY RELATION} 

\author{Lorraine A. Allen and Margaret J. Geller}
\affil{Harvard-Smithsonian Center for Astrophysics} 
\authoraddr{60 Garden Street, Cambridge, MA 02138}  
 
\begin {abstract}

We have assembled 21-cm linewidths
for 376 of  
the 732 galaxies in a magnitude-limited redshift survey of the Perseus-Pisces
region. 

We analyze a distance limited subset of 271 galaxies (131 widths) to
examine the relationship between linewidth and local density. The
sample indicates that galaxies with linewidths $\gtrsim 480$ km/s
are absent from regions where the galaxy density is $\lesssim 0.03$
galaxies Mpc$^{-3}$ ($M_{B(0)}< -18.3$).  This effect is in the direction
predicted by standard CDM models. 
Galaxies with
linewidths $\lesssim$ 480 km/s appear throughout
the sample.

The observational constraints 
could be susbtantially improved with a larger sample, IR photometry, and
more uniform 21-cm data.
  
\end{abstract}
 
\section{Introduction}

Cold Dark Matter (CDM) models are  the most successful in accounting
for the observed properties of large scale structure
in the universe (see Ostriker 1993 for a review; Zurek {\it et al.} 1994,
Ueda {\it et al.} 1994).  Biased CDM models, in which galaxies
are more clustered than the overall mass distribution, can account for the
observed galaxy clustering on scales $\lesssim 10 h^{-1}$ Mpc, rms peculiar
velocities of galaxies, and the qualitative appearance of large scale
features - voids, filaments, and walls (White {\it et al.} 1987b, 
Park 1990, Cen and 
Ostriker 1992). 
In addition, these models descibe many of the properties
of individual galaxies, including dark haloes around spirals with
flat rotation curves; they also provide an explanation for the formation
of the different morphological types and for the observed
morphology-density
relation (Cen and Ostriker 1993, Frenk {\it et al.} 1985).  

CDM models with ${\Omega}h$ = 1/2 
fail to account for the observed galaxy distribution  on large
scales $\gtrsim 30 h^{-1}$ Mpc (Park {\it et al.} 1994, Ueda {\it et al.}
1993, Gelb and Bertschinger 1994). 
The models also fail to account for  observed large scale flows
(White {\it et al.} 1987b, Strauss {\it et al.} 1995). Minimally
biased low-density CDM models
with ${\Omega}h \simeq$ 0.2 -- 0.3  (with and without a non-zero cosmological
constant) are consistent with
the large-scale galaxy distribution and with the amplitude of mass
fluctuations at the current epoch implied by COBE
(Kofman {\it et al.} 1993, Ueda {\it et al.} 1993,
Park {\it et al.} 1994, 
Cen and Ostriker 1994).
 
Galaxy formation in biased CDM models occurs preferentially in regions
where the
density exceeds a threshold. In all
CDM models, the most massive galaxies tend to form  
in the  densest regions; here the matter collapses and accretes more
rapidly than at  lower density.
Furthermore,
high density regions cluster (thus galaxies are
more clustered than the overall mass distribution)  
and thus the most
massive galaxies tend to be more clustered.
Simulations of CDM models demonstrate these effects (White {\it et al.} 1987a,
Cen and Ostriker 1993, Evrard {\it et al.} 1994).

Because mass and luminosity are correlated, the models predict luminosity
segregation. In other words, the brightest galaxies should be preferentially
in the
densest regions.
Park {\it et al.} (1994) observed this effect in the CfA survey
at about the 2$\sigma$ level.  
Davis {\it et al.} (1988) found, based on correlation functions for the
CfA1 and Southern Sky surveys, that brighter galaxies cluster more
than fainter ones.  
Similarly, Gramann and Einasto (1992) claim 
that bright galaxies   
preferentially populate
high-density regions in the Virgo, Coma, and Perseus superclusters
on scales 3 -- 20 $h^{-1}$ Mpc.

Rotation velocity is a more direct and cleaner indicator of mass than
is the luminosity; thus, it should also be a function of local galaxy density. 
The rotation velocity $\Delta V$ and the luminosity, L,  obey
the Tully-Fisher relation,
$L \propto \Delta
V^{\beta}$ where 
$\beta = 2 - 4$
depending on color; $\beta \simeq 2$ at B with large scatter.
The scatter at B results  largely
from 
internal extinction in the galaxies
(Aaronson {\it et al.} 1979).  Examining the correlation
between rotation velocity and local galaxy density is a more direct
constraint on CDM models than a relationship between luminosity and
local density.

So far, White {\it et al.} (1988) have made the
only attempt to assess the rotation velocity-local density relation.
They analyzed a sample of 1387 galaxies from the
NGC and found a correlation
between rotation velocity
and density consistent with CDM.  However, only about one third of the
galaxies have  measured linewidths; most of the widths
are actually derived from Tully-Fisher relations based on B magnitudes from
the NGC.  The large scatter in the Tully-Fisher relation at B diminishes the
impact of using the rotation velocities. Their sample is
also affected  by incompleteness and by the relatively large peculiar
motions within the Local Supercluster.

Here we examine the relationship between rotation velocity and local
galaxy density for a complete sample of 376 galaxies
with measured HI linewidths; the sample is in the
Perseus-Pisces supercluster region and is deep enough that
Virgo infall is not an issue. 

Section 2 describes the dataset and tests for systematic bias and
incompleteness. In Section 3 we use a nearest neighbor statistic to evaluate
the relationship between linewidth and local galaxy density. We conclude
in Section 4 and discuss the prospects for improved constraints.

\section{The Data}

We have assembled 21-cm linewidth measurements
(Giovanelli and Haynes 1985 (PP1), Giovanelli {\it et al.} 1986 (PP2),
Giovanelli and Haynes 1989 (PP4), Giovanelli and Haynes 1993 (PP6),
Wegner {\it et al.} 1993 (PP5), Haynes and Giovanelli 1984 (IG4),
Huchtmeier and Richter 1989, Eder {\it et al.} 1991 (EGH),
Scodeggio and Gavazzi 1993 (SG), Lu {\it et al.} 1988 (LHGRL)) for 376 of  
the 732 galaxies in a magnitude-limited redshift survey of the Perseus-Pisces
region.  Here we describe the dataset and examine it for systematic
biases.

\subsection {Linewidths for a Redshift Survey}

The Center for Astrophysics redshift survey (Geller and Huchra 1989,
Huchra {\it et al.} 1990, Huchra {\it et al.} 1983, Vogeley 1993)
covers large regions of the northern and southern galactic hemispheres
to a limiting magnitude m$_{Zw} = 15.5$ in the
Zwicky (1961-1968) catalog.  Here we consider the subsample of
732 galaxies in the southern
galactic hemisphere with $cz \leq$ 8500 km s$^{-1}$ and
with right ascension $22^h < \alpha < 3^h$ and
declination $15^{\circ} \leq \delta < 33^{\circ}$ (the MLS hereafter).
This region minimizes the effect of Galactic obscuration and
maximizes the completeness of the sample of galaxies with known linewidth.

We assembled 21-cm linewidths for 376 of the galaxies in the
redshift survey (the MLWS hereafter).  Giovanelli
and Haynes (PP1, PP2, PP4, PP5, PP6, IG4) measured 370 of these linewidths
at Arecibo.  The other 6 linewidths are from the literature
(EGH, SG)
or from the
Huchtmeier-Richter Catalog (1989).  372 of the
measurements are 50\% linewidths,
2 are 20\% measurements, and 2 measurements are averages of the 50\% and 20\%
widths.  All linewidths are corrected for inclination and for
redshift broadening.  We include only galaxies with inclination
$i \geq 40^\circ$.  Among the galaxies with measured linewidths, 288 are
spirals, 39 are S0 or S0/a, and 49 are irregular, peculiar, or
unclassifiable.  

HI profiles for $142$ of the 376 galaxies are published
(PP1, PP2, IG4, EGH, LHGRL, SG, Haynes 1981, Hewitt {\it et al.} 1983).
Of these,
$105$ profiles are clean, steep-sided profiles
with flat baselines. 
The remaining profiles, because of shallowness or
curved baselines, may
yield linewidths with larger errors.
However, because we do not have a complete set of profiles, we use
all of the data. 

\subsection {Tests for Incompleteness and Systematic Bias}

The MLS is a complete magnitude limited survey.  Here we examine
the completeness of the MLWS and we test for biases in the
magnitude and redshift distributions relative to those for the MLS.

For the MLS sample as a whole, we do not have morphological types. However,
all 211 MLS galaxies with m$_{Zw}\leq 14.5$ have types 
(Huchra {\it et al.} 1983).    
Of these, 113 galaxies are spirals; 76 of these spirals, $36\%$ of the
entire 211 galaxy sample, have $i\ge 40^\circ$ and are contained within
the MLWS.
Assuming that the morphological composition
and orientation of galaxies of the samples are  
independent of the magnitude limit, the MLS should
likewise contain $\sim 36\%$, or 264, spirals with $i\ge 40^\circ$;
the MLWS contains 288 spirals, $39\%$ of the MLS sample, in
reasonable agreement with expectation.

If galaxies were randomly oriented we would expect to find
42\% of spirals with $i\geq 40^\circ$. Because galaxies are apparently
fainter when they are more edge-on, the fraction of edge-on galaxies
in a magnitude limited sample should be systematically low. The fractions
of edge-on spirals are systematically low in both the MLS and the MLWS, but
the bias is $\lesssim 2\sigma$ where $\sigma$ is the $\sqrt{N}$
error in the count.  We do not model this bias because the error in the
model is likely to exceed the amplitude of the bias.

Figure 1 shows the distribution of MLWS galaxies within the
MLS survey in two $9^\circ$ declination slices. Because of
the well-known morphology-density relation, (Dressler 1980, Postman and Geller
1984), the MLWS galaxies are less common in the cores of
clusters than in lower density regions.
Figures 2a and 2b show the ratio of MLWS to MLS galaxies as a function
of apparent magnitude and redshift, respectively. There is no apparent bias
as a function of apparent magnitude;
in figure 2a all galaxies with $m_{Zw} \leq 13.0$ are in the
first bin.  
The distribution as a
function of redshift again reflects the morphology-density relation. In
the low density void at $cz \lesssim  
$4000 km s$^{-1}$, the MLWS fraction
is somewhat larger than in the Perseus Pisces
supercluster at $cz \simeq$ 5000 km s$^{-1}$.

From this set of comparisons
we conclude that the MLWS is remarkably free of systematic biases
relative to the MLS. The MLS itself may contain biases, but unless these
biases are correlated in some way with linewidth, they should not affect
the analyses here.

Figure 3 shows the Tully-Fisher (TF) relation for the sample.
The squares are the quartile of MLWS
galaxies with the largest widths; 
the triangles are the lowest quartile. 
The lowest linewidth galaxies are especially faint; 
for these objects the Tully-Fisher relation is not linear  
(see Mould {\it et al.} 1989 and references therein).
Aaronson {\it et al.} (1986) found that a quadratic form of the relation
(in the IR) fit their data well.  This non-linearity of the
TF relation, along with increased scatter in the blue band TF
relation, may account for the data
in figure 3. 

Because the smallest linewidth galaxies are so faint,  
they are naturally missing from the MLWS at larger redshifts.  The brightest
galaxies with the largest linewidths are visible
throughout the sample region. 

\section {The Correlation of Linewidth with Local Density} 

Here we examine the correlation of linewidth with local galaxy
density. 
To avoid
binning and thus make use of each linewidth measurement
as an
independent estimator of the linewidth-density relation, we
compute the local density by identifying the N nearest neighbors
of each MLWS galaxy.
Dressler (1980) and
Dressler and Shectman (1988) used this method to examine the 
morphology-density relation and to investigate velocity substructure
within clusters of galaxies.  

\subsection {The Nearest Neighbor Density Estimator} 

The density in the neighborhood of the MLWS galaxy is 
\begin{equation}
{\rho/\rho_{avg}}= {N\over
{{8\over3}\pi D^3 {\int_{-\infty}^{M_g}\Phi(M)dM}}}   
\end{equation}
where D is the median projected distance to the N (=10) nearest     
neighbors; N=10 provides a stable estimate of the density for the
MLWS galaxies. 
$\Phi(M)$ is the luminosity function for the sample and $M_{g}$ is
\begin{equation}
M_g =  m_{lim} -25 -5\log(V_g/H_\circ)
\end{equation}
where $m_{lim}$ is the magnitude limit and $V_g$ is the 
MLWS galaxy velocity. 
The integral in the
denominator corrects the galaxy density
for the unseen portion of the luminosity function
in a magnitude limited sample.  

We compute the projected separation between galaxies:
\begin{equation}
D = \sin(\theta/2)(V_g+V_i)/H_\circ,  
\end{equation} 
where 
$V_g$ and $V_i$ are the velocities of the two galaxies in
the pair (the subscript g refers to the central MLWS galaxy),
$\theta$ is their angular separation, and $H_\circ$ is the Hubble
constant.  We take $H_\circ = 100$ km/s/Mpc throughout.
We also require that the velocity difference between the MLWS galaxy
and each of its nearest neighbors is less than 
a cutoff value, $V_L$.  To account for the magnitude limiting,
we scale $V_L$ as  
\begin{equation} 
V_L = V_\circ\left[\int_{-\infty}^{M_{lim}} \Phi(M)
{dM}/\int_{-\infty}^{M_g} \Phi(M) {dM}\right]^{1/3},
\end{equation}
(Huchra \& Geller 1982)
where $M_{lim}$
is the faintest absolute magnitude at which galaxies in a sample with
magnitude limit $m_{lim}$ are visible at fiducial velocity $V_F.$
$M_{lim}$ is then  
\begin{equation}  
M_{lim} = m_{lim} -25 -5\log(V_F/H_\circ)  
\end{equation} 
and $M_g$ is given by Eq.(2).   
$V_\circ$ in Eq.(4) is 1000 km/s
with $V_F = 8000$ km/s (Eq. (5)).  The resulting values of $V_L$
prevent clipping of the velocity dispersion of rich clusters
but they are not so large that we cross the $\sim$ 5000 km/s voids
in the survey. 
Varying V$_\circ$ by $\sim$30\% produces a negligible effect on the local
densities.

For a distance-limited sample, we compute  
the local density 
\begin{equation}
{\rho/\rho_{avg}}= {N\over
{{8\over3}\pi D^3 {\int_{-\infty}^{M_{lim}}\phi(M)dM}}}
\end{equation}
where $M_{lim}$ is given by Eq.(5) with $V_F = 1000$ km/s.
The corresponding $V_L$ is constant at 800 km/s, comparable
with the typical velocity
dispersion of a cluster.  Again the density estimates are
insensitive to changes in $V_L$. In contrast with Eq.(1) for the magnitude
limited sample, the integral in the denominator is constant here.  

Figure 4a shows the correlation between linewidth and density for the
nearest neighbor   
estimator applied to the magnitude limited MLWS.
Squares are the highest density quartile; triangles are the
lowest.  Spearman's rank correlation coefficient is 0.210, a
$3.9 \times 10^{-3} \%$ probability of no correlation.  
Galaxies with linewidths $\gtrsim 480$ km/s
preferentially occupy denser regions: there is an absence
of these galaxies in the least dense regions in figure 4a.  

At the largest densities, the most massive
galaxies are early type. The absence of these galaxies from the sample
with measured linewidths actually surpresses the correlation. This
effect is most evident for $\rho/\rho_{avg} > 2$ and
$\Delta V > 620$ km s$^{-1}$.  
                                                              
Figure 5a shows the distribution of linewidths for the MLWS galaxies
with surrounding density in the lowest and highest quartiles.
There is a clear shift of the linewidth distribution toward larger
linewidth at larger density. The median linewidth is $316^{+92}_{-77}$
km/s for the lower quartile and $400^{+72}_{-89}$ km/s for the upper
quartile.  The errors are the interquartile range for each distribution.  
A K-S test on the unbinned data for figure 5a shows that there is 
 $<0.04 \%$ chance that they are drawn from the same 
distribution.

The behavior of the galaxies with the smallest linewidths is hard
to assess from this sample.   
The void region at $cz \lesssim$ 4000 km/s and 
the Perseus Pisces supercluster at 
$cz \simeq $ 5000 km/s introduce an artificial 
correlation between redshift and 
density. (The Spearman rank coefficient is 0.248).   
The Tully-Fisher relation (figure 3) implies that galaxies 
with small linewidths
tend to be faint.  They are therefore not visible  
in our magnitude-limited sample at the 
higher redshifts where the sample is denser. 

To remove the artificial correlation between redshift and density
and thus
to investigate the distribution of small linewidth 
galaxies with local density,
we construct a distance-limited sample, ABS1, with  
velocity $cz \leq
5700$ km/s and $M_{B(0)} \leq -18.3$. 
We make this choice to maximize the number of galaxies with known widths: we
have 271 galaxies; 131 have measured widths.    
Because this sample is  
more sparse than the magnitude-limited MLWS,
we choose the number of nearest neighbors for the ABS1 sample which best
approximates the density estimates for the entire MLWS sample.  The
choice N=8 satisfies this criterion.

Figure 4b shows the results for ABS1.
The squares are the highest density quartile; the triangles are
the lowest quartile.   
The $r_s$ value is 0.117, an $18.4 \%$ probability of no correlation.
The galaxies with large linewidths ($\gtrsim 480$ km/s)  
are absent from regions with log ($\rho/\rho_{avg}) \lesssim -0.4$.       
There is no such effect for galaxies with
small linewidths ($\lesssim 350$ km/s).  

Figure 5b shows the distribution
of linewidths for galaxies in ABS1 with surrounding densities
in the lowest and highest
quartiles.  
The linewidth distribution again shifts toward
larger linewidth for galaxies in regions
of larger local density.  The median linewidth for galaxies in regions
of smallest and largest densities are, respectively, $340^{+99}_{-60}$
and $411^{+72}_{-79}$ km/s, where 
the quoted error is the interquartile range.
The difference between the two linewidth distributions for the highest
and lowest density regions is significant at roughly the
1.2$\sigma$ level; a K-S test on the unbinned 
data for figure 5b shows that there is  $14.3 \%$ probability that
they are drawn from the same distribution. This estimate of
significance is probably conservative; figure 4b shows that the first
quartile contains some galaxies 
with large linewidth. The small sample size
and the dangers of {\it a posteriori} statistics
prevent us from further dividing the sample, which would probably
increase the significance.

Figures 6a and 
6b provide strong qualitative support for the absence of galaxies with large
linewidth from low density regions: the figures show the
distribution in redshift space of the quartile of ABS1 galaxies 
with the largest
and smallest widths, respectively.   
The 
squares and triangles are the largest and smallest ABS1 quartile galaxies,
respectively, 
and the crosses 
are the remaining galaxies in
the sample (with and without widths).
Figure 6a shows that the
galaxies with the largest linewidths occupy the densest regions and are 
absent from
the voids. However (Figure 6b), the smallest linewidth  
galaxies appear in the voids as well as in higher density regions.

In summary, galaxies in the MLWS
of all linewidths populate regions with 
$\log \rho/
\rho_{avg} \gtrsim 0.5$ ($\rho \gtrsim 0.18$ $Mpc^{-3}$).    
However, underdense regions, with $\log \rho/\rho_{avg} \lesssim 0.1$
($\rho \lesssim 0.07$ $Mpc^{-3}$),  
are deficient in  
galaxies with linewidths $\gtrsim 480$ km/s.

The effect we find is 
analogous to the luminosity segregation observed by Park  
{\it et al.} (1994) in 
which galaxies of all luminosities occupy moderate and high 
density regions; the brightest galaxies are absent from the lowest density
regions.   
For local densities $\rho < 0.015 h^3 Mpc^{-3}$ the fraction
of bright galaxies is $\simeq 2\sigma$ smaller than expected for a
random sample;  the bright galaxy fraction is consistent
with random for densities $\rho > 0.03 h^3 Mpc^{-3}$ $(M_{B(0)}<-19.1).$ 

As a consistency check for our ABS1 sample, we plot $M_{abs}$
as a function of density, $\rho,$ in figure 7  
(compare figure 12 in Park {\it et al.} (1994)).
Open squares represent the quartile with the largest widths;
solid triangles are the lowest quartile.  
At the largest densities, the brightest galaxies are
generally absent because early type galaxies are not included
in the sample of galaxies with measured linewidths. A subtle
luminosity segregation appears  in our sample at densities
$\lesssim 0.03 h^{3} Mpc^{-3}$; 
the effect is less pronounced than  in Park {\it et al.} (1994) because
of our smaller  sample size.  This 
luminosity segregation is a natural consequence of  
the dependence of the Tully-Fisher relation
(figure 3).  However, because of the large scatter in the
TF relation (especially at small linewidth), the direct
rotation velocity - density relation is the physical relation
we seek.

\section{Discussion}

Our analysis of a distance limited sample of 271 galaxies
(131 have measured linewidths) provides evidence for a
dependence of linewidth on local galaxy density. The
sample indicates that galaxies with linewidths $\gtrsim 480$ km/s
are absent from regions where the galaxy density is $\lesssim 0.03$
galaxies Mpc$^{-3}$ ($M_{B(0)}<-18.3$). This effect is in the direction
predicted by CDM models. 

The analysis of this small sample suggests that the linewidth-local
density relation could provide interesting constraints on models
for galaxy formation and large-scale structure. The observational
constraints could easily be improved by  increasing the sample size.
Figures 4 and 5 show that even doubling the sample
size would provide far more convincing evidence of the absence
of galaxies with large linewidth in the lowest density regions.

An important limitation of our analysis is the use of B magnitudes
from Zwicky (1961-1968). These data are not of uniform quality and
the Tully-Fisher relation at B is broad. A much more powerful test of the
linewidth-local density relation could be made with uniform IR
photometry. An unbiased set of steep-sided 21cm profiles at high
signal-to-noise would also contribute to tighter constraints.
On the theoretical side, it would be useful to compute the
behavior of the rotation velocity-density relation in terms of the
observable parameters.

We
thank Riccardo Giovanelli for providing data in a computer
readable form. We also thank
Antonaldo Diaferio, Emilio Falco, and Michael Kurtz
for detailed comments on the manuscript.
Lorraine Allen was partially supported by a NASA Graduate
Fellowship. This research was supported in part by NASA Grant NAGW-201
and by the Smithsonian Institution.

\newpage  

\centerline {\bf Figure Captions}
 
\noindent {\bf FIG. 1.} The distribution of MLWS galaxies (squares) among
the MLS galaxies (crosses) in two $9^\circ$ declination slices.
 
\noindent {\bf FIG. 2.} {\bf a)} The ratio of
MLWS to MLS galaxies as a function of
apparent magnitude.  All galaxies with $m_{Zw} \le 13.0$ are in the first
bin. {\bf b)} The ratio of MLWS to MLS galaxies as a function of redshift.
The parentheses contain the number of MLWS/MLS galaxies.
 
\noindent {\bf FIG. 3.} The Tully-Fisher relation for the MLWS.  Squares
are the quartile of the MLWS with the largest widths; triangles are
the lowest quartile.
 
\noindent {\bf FIG. 4.} {\bf a)} Linewidth as a function of density
($\rho/\rho_{avg}$), computed using the Nearest Neighbor Density
estimator (N=10), for the MLWS.  Squares are the quartile of the MLWS
with the largest widths; triangles are the lowest quartile.
{\bf b)} Linewidth as a function of density
for the ABS1 galaxies. N=8 for the density estimator.  Squares and
triangles are the quartiles of ABS1 galaxies with the largest and
smallest widths, respectively.
 
\noindent {\bf FIG. 5.} {\bf a)} The distribution of linewidths for galaxies
in the MLWS with densities in the lowest and highest quartiles
of the density distribution.  The median linewidth for the lower
and upper linewidth distributions are, respectively, $316^{+92}_{-77}$
and $400^{+72}_{-89},$ where the quoted error is the interquartile range.
{\bf b)} The distribution of linewidths for ABS1 galaxies
with densities in the lowest and highest quartiles
of the density distribution.
The median linewidth for the lower
and upper distributions are, respectively, $340^{+99}_{-60}$
and $411^{+72}_{-79}.$
 
\noindent {\bf FIG. 6a.} The distribution in redshift space of the
quartile of ABS1 galaxies with the largest linewidths (squares)
among the remaining galaxies (with and without widths) (crosses).
 
\noindent {\bf FIG. 6b.} The distribution in redshift space of the
quartile of ABS1 galaxies with the smallest linewidths (triangles)
among the remaining galaxies (with and without widths) (crosses).
 
\noindent {\bf FIG. 7.} Absolute (blue) magnitude as a function of
density, $\rho,$ for the ABS1 sample.  Open squares are the quartile
of galaxies with the largest linewidths; solid triangles are the
lowest quartile.
 
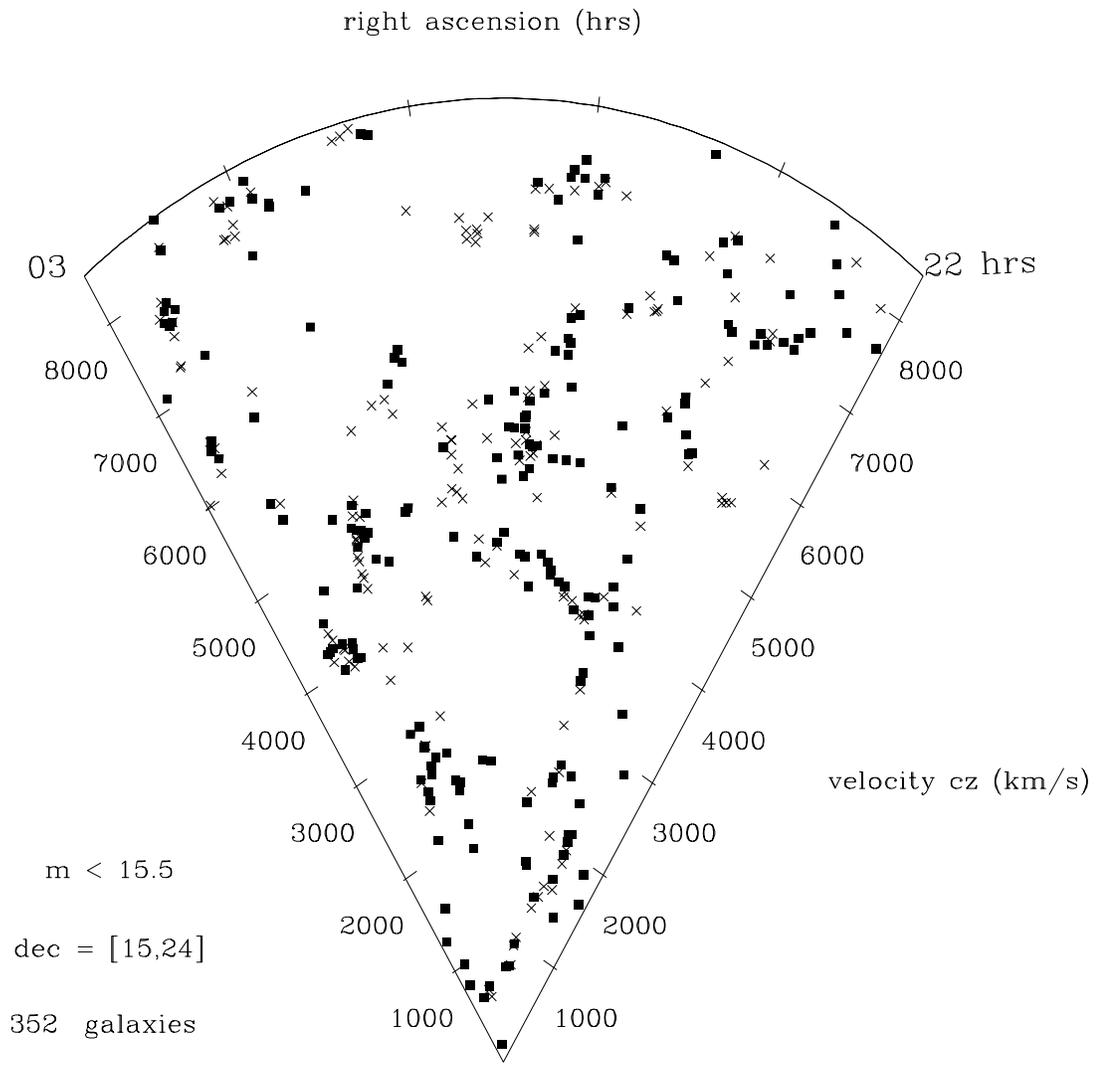
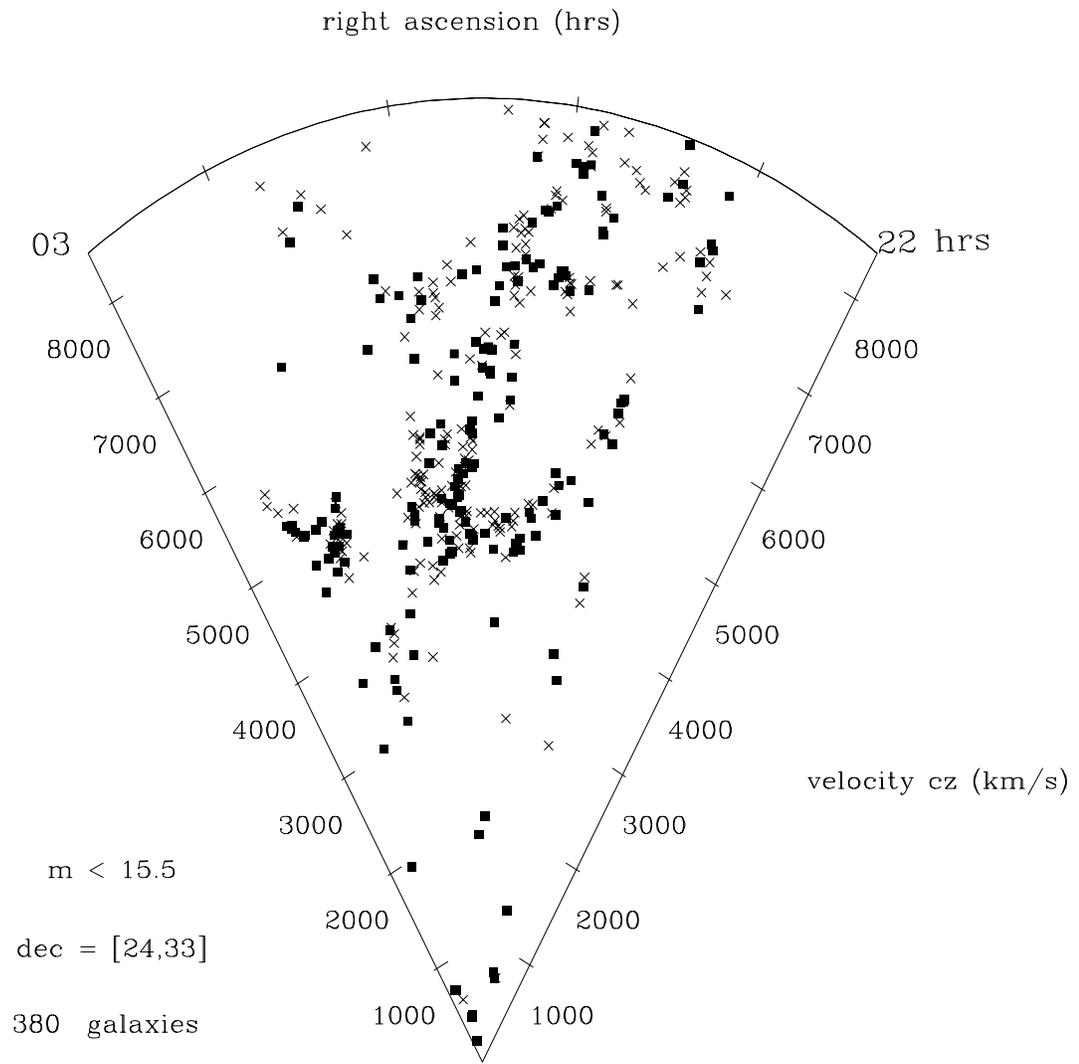
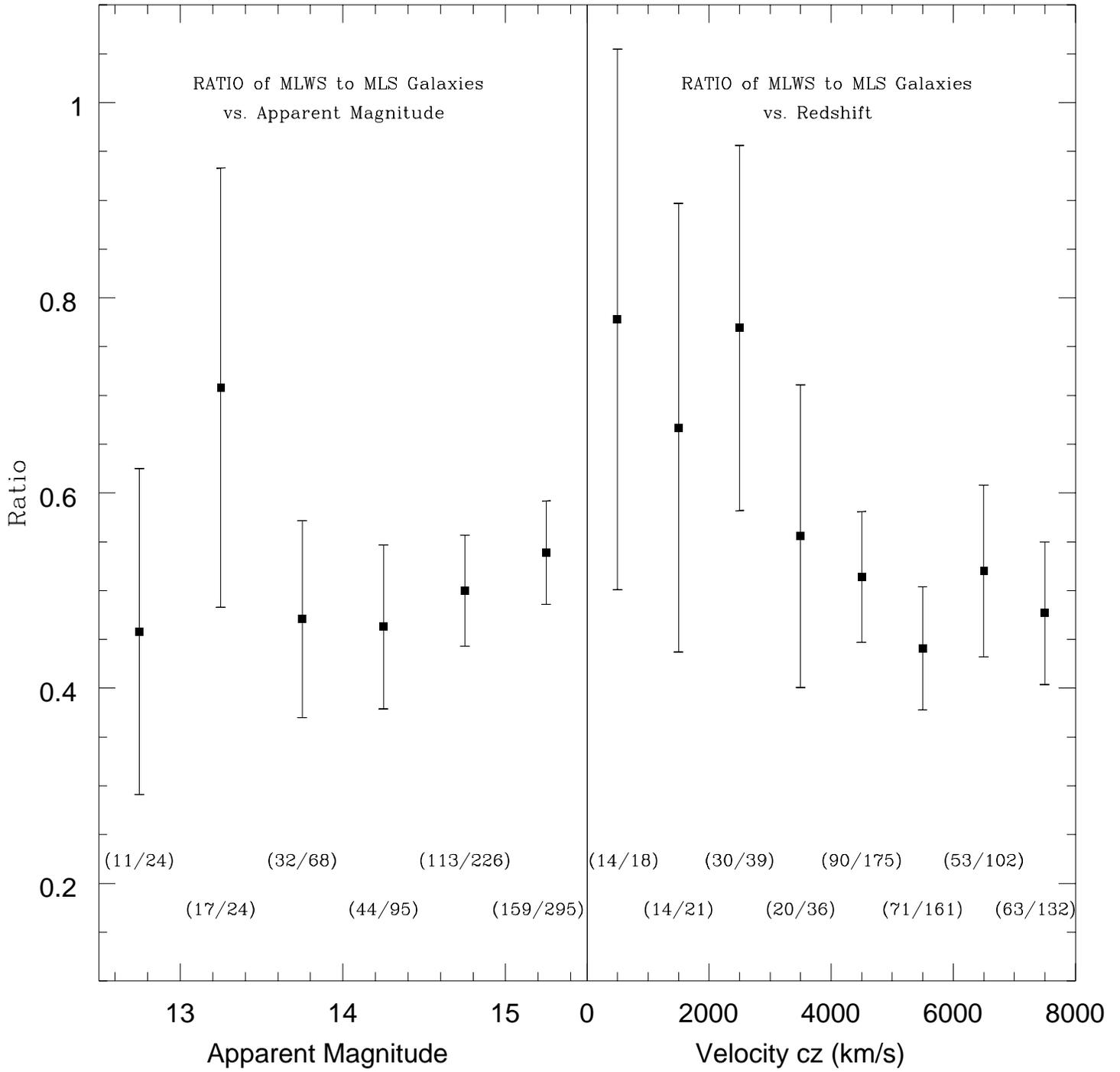
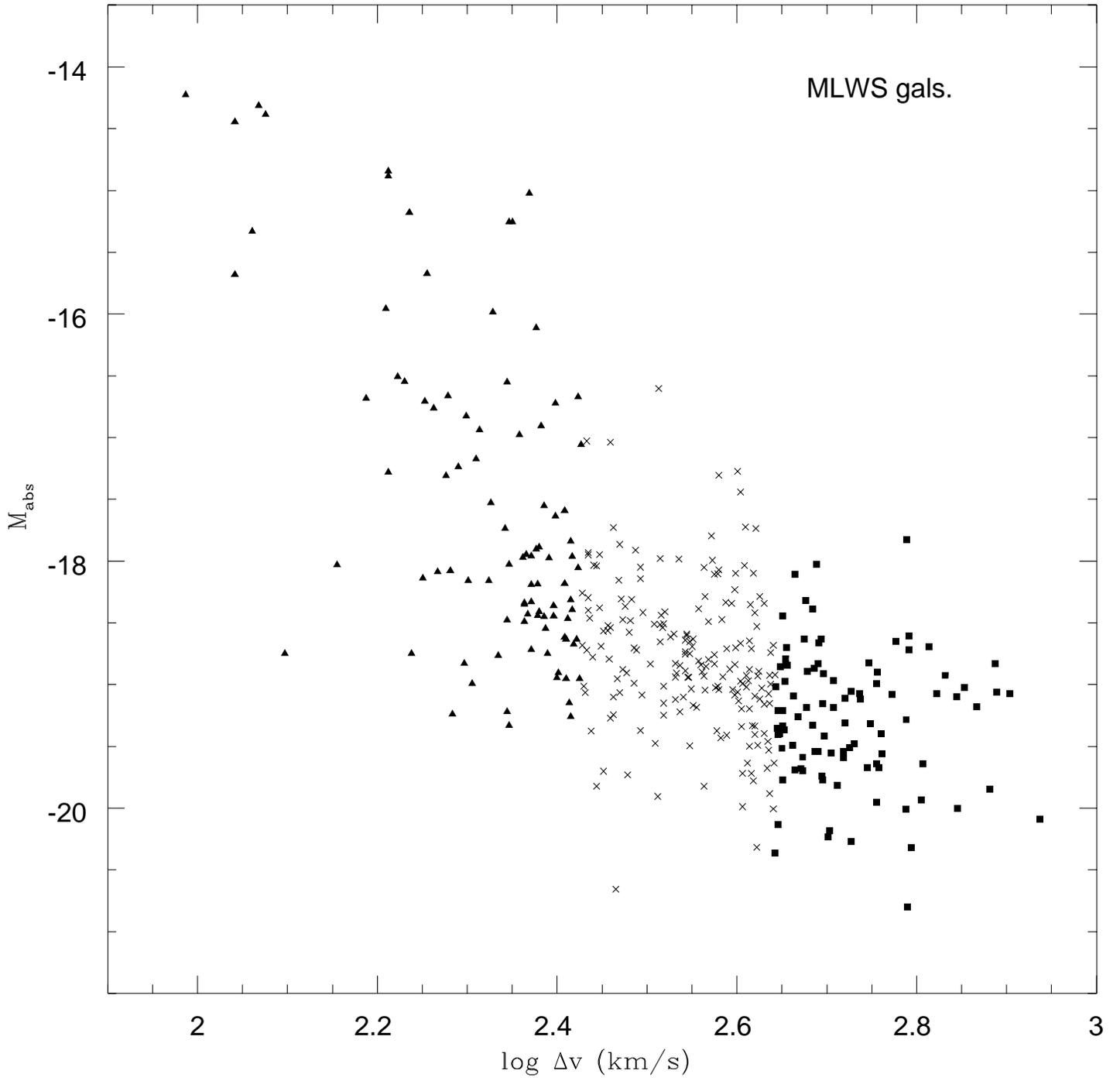
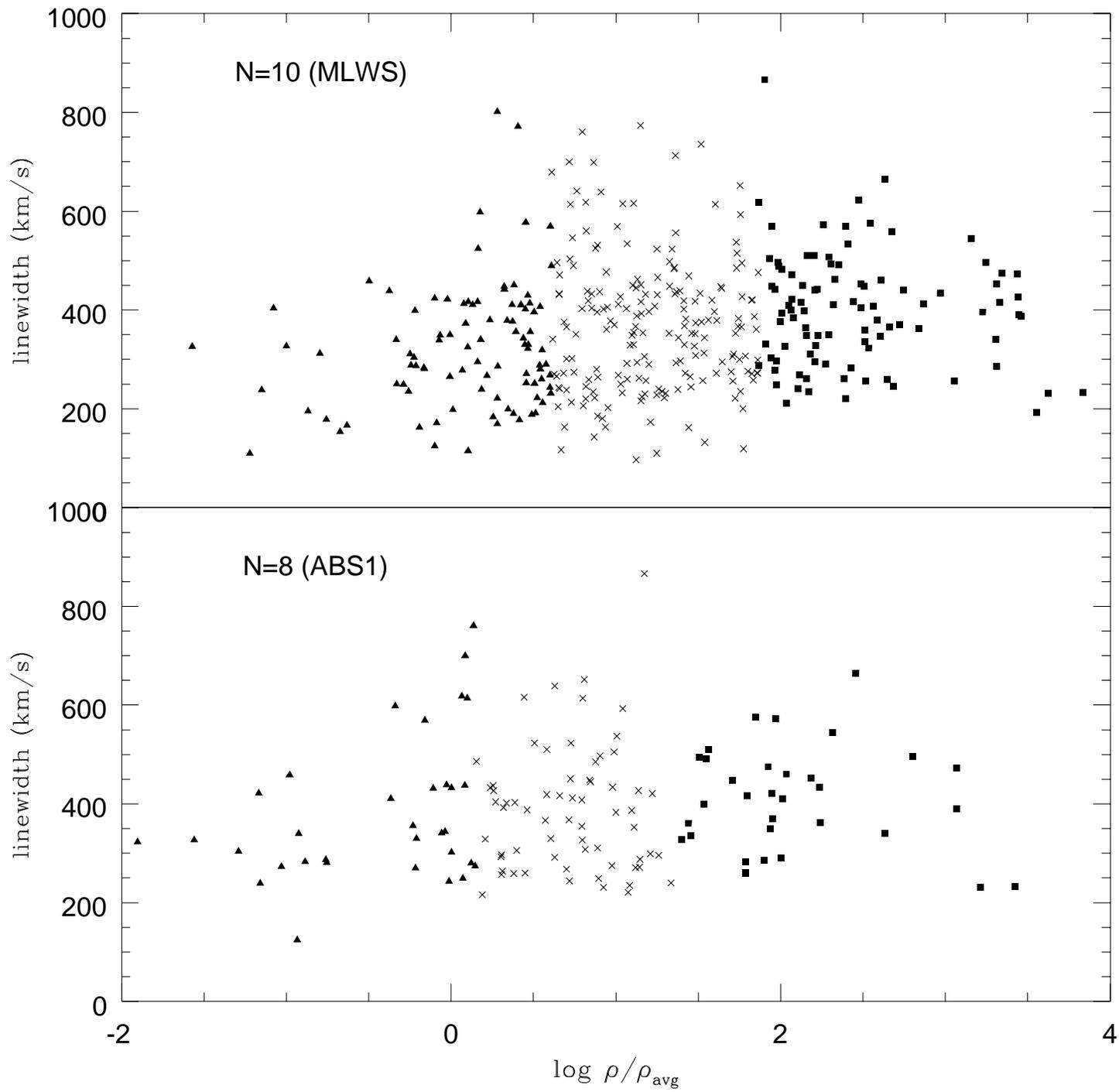
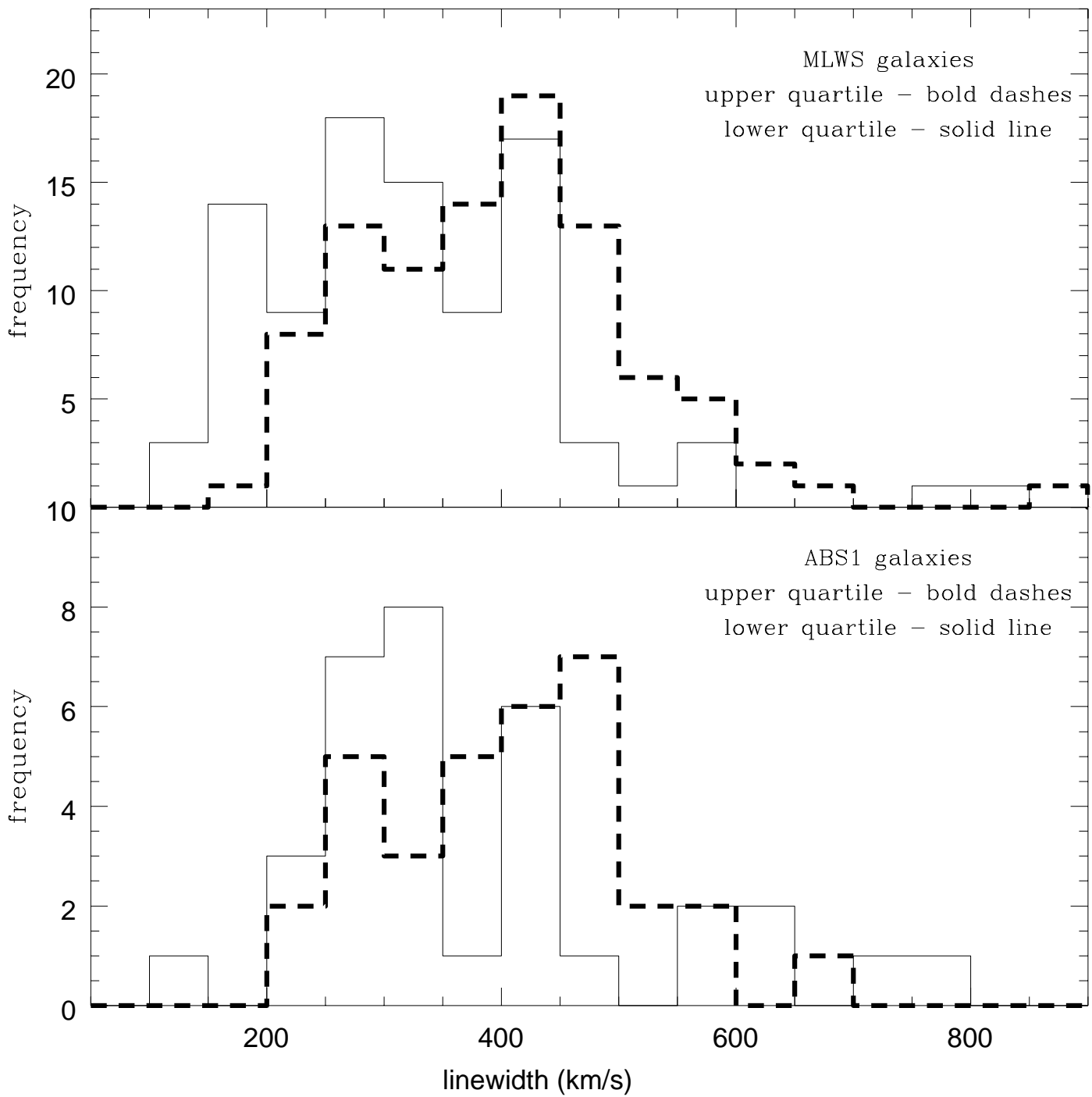
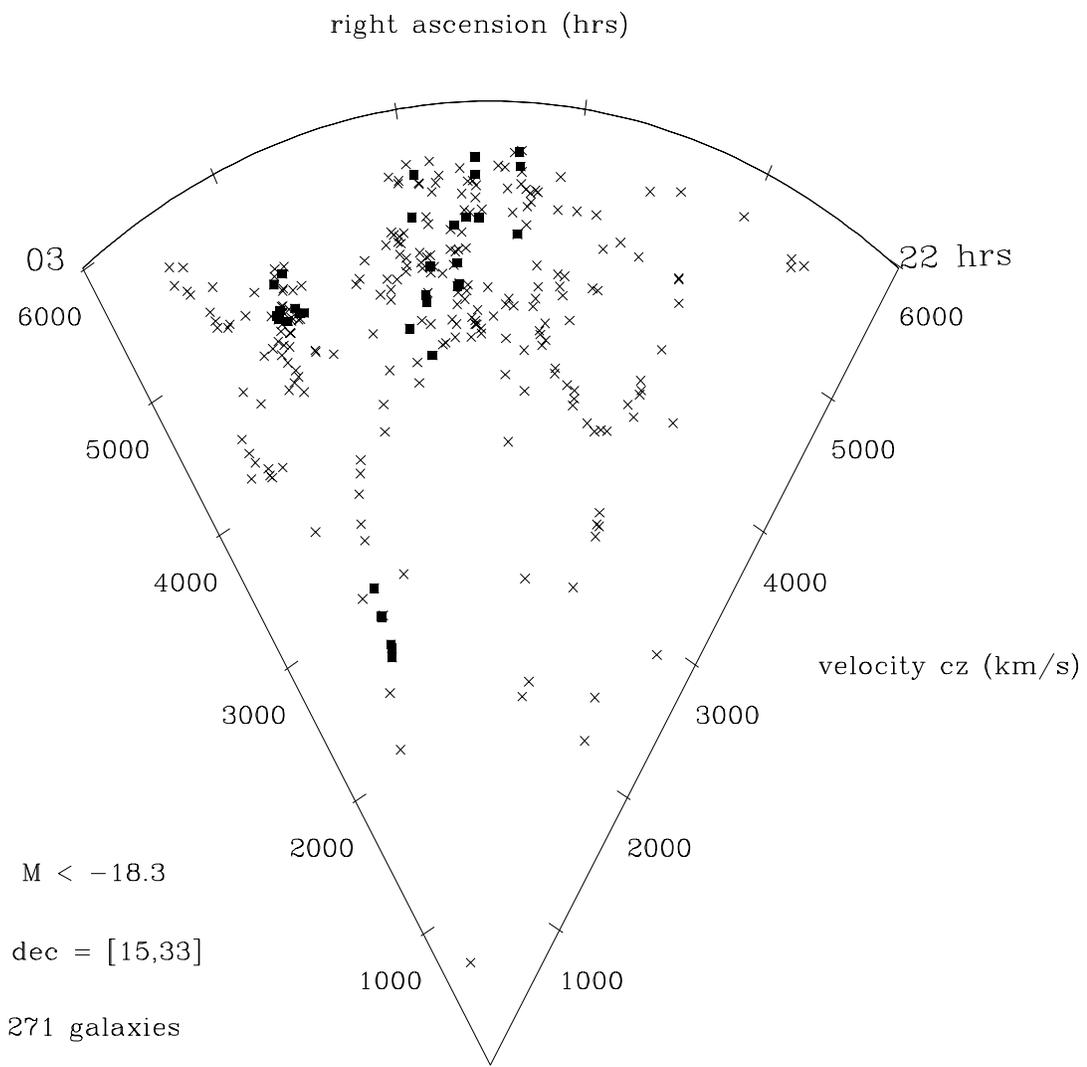
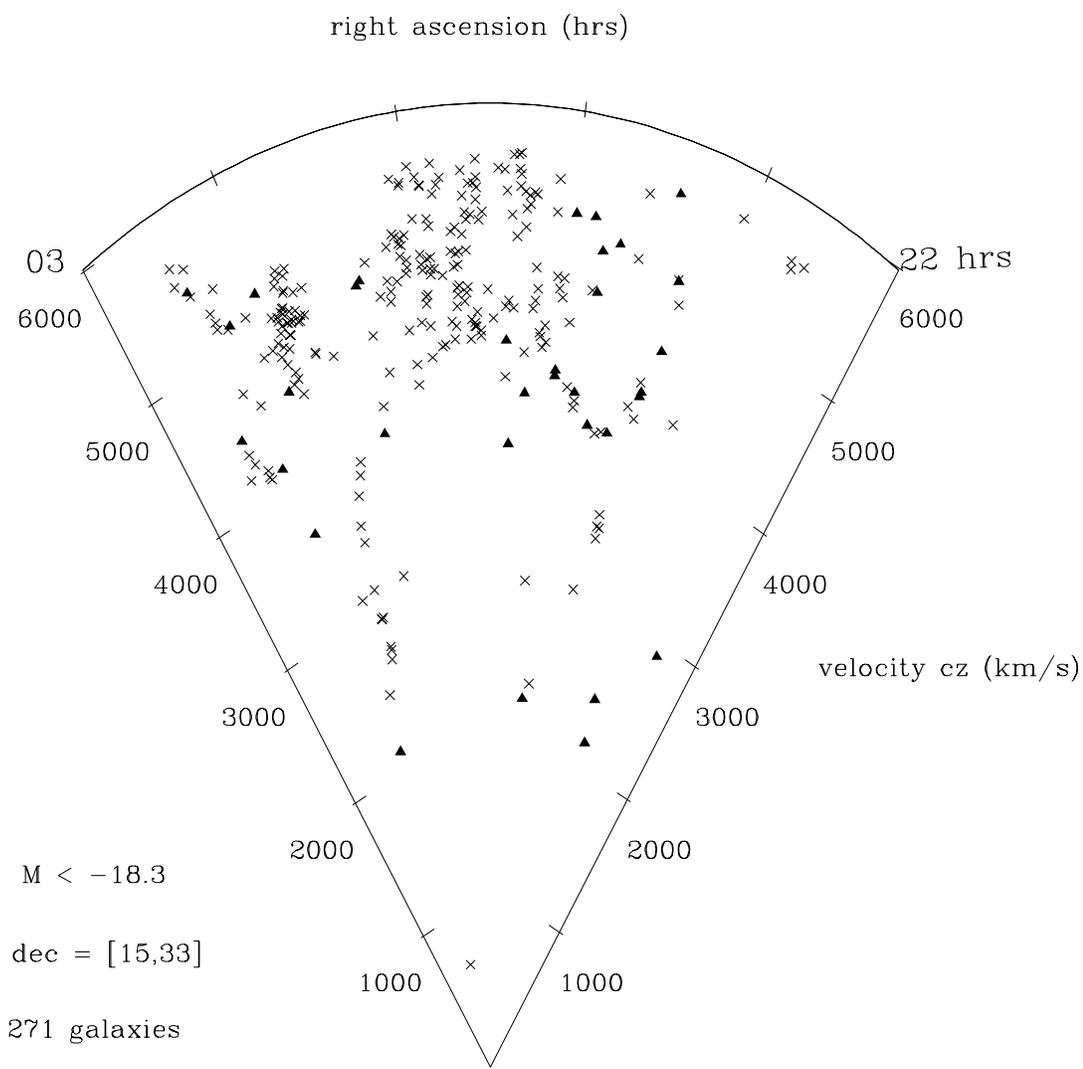
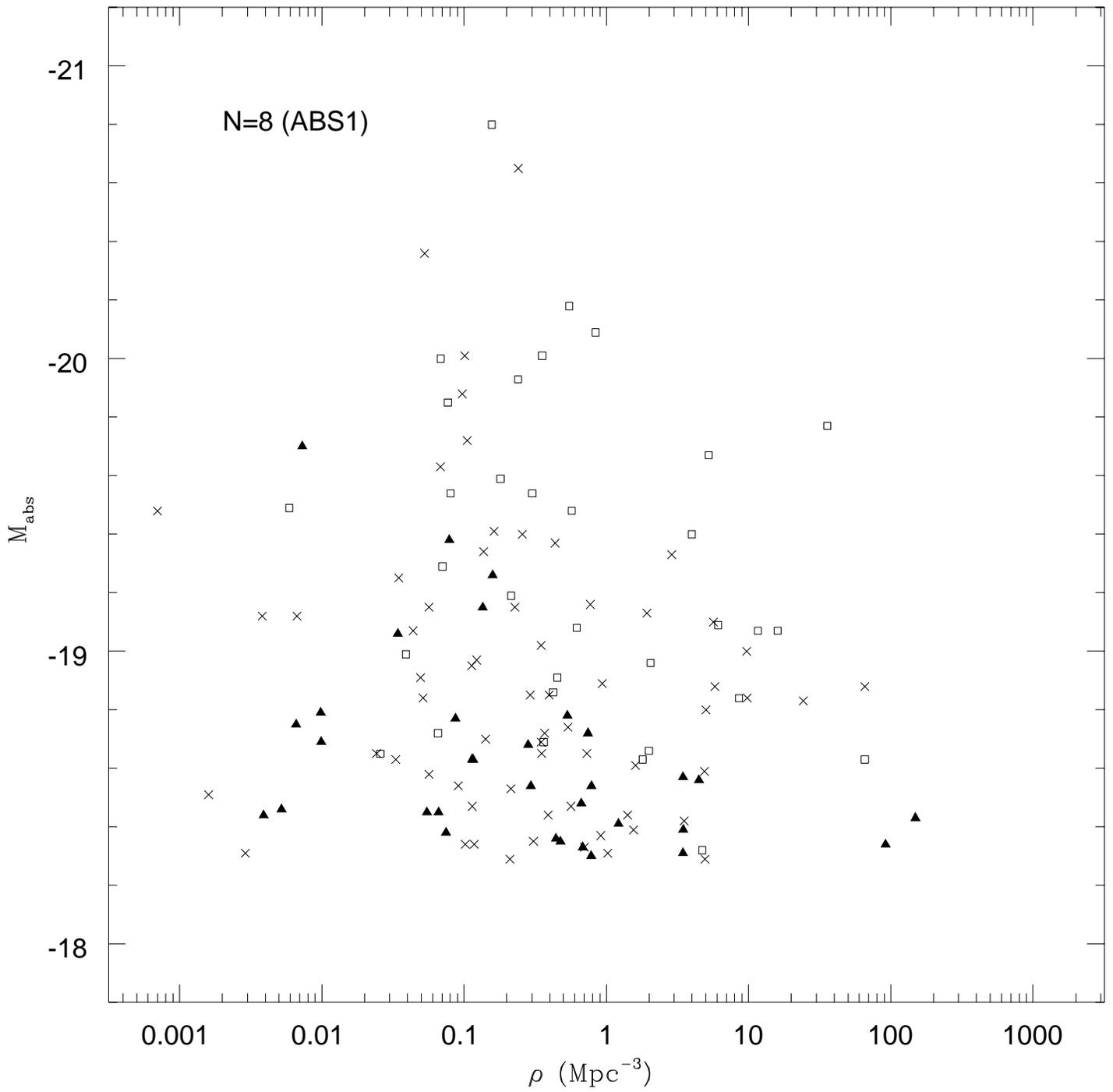
\end{document}